\documentclass[runningheads]{llncs}
\usepackage[T1]{fontenc}
\usepackage{graphicx}
\usepackage{longtable}

\usepackage{array}

\begin{document}
\title{Attention and Sensory Processing in Augmented Reality: Empowering ADHD population
}
\titlerunning{Attention in Augmented Reality}
%
\author{Shiva Ghasemi\inst{1}\orcidID{0000-0003-1435-6342} \and
Majid Behravan\inst{1}\orcidID{0000-0001-6525-6646} \and
Sunday Ubur\inst{1}\orcidID{0009-0004-2172-7003} \and
Denis Gra{\v{c}}anin\inst{1}\orcidID{0000-0001-6831-2818}}
\authorrunning{S. Ghasemi et al.}
%
\institute{Virginia Tech, Blacksburg, VA 24060, USA 
\\
\email{\{shivagh,behravan,uburs,gracanin\}@vt.edu}}

\maketitle              
\begin{abstract}

The brain's attention system is a complex and adaptive network of brain regions that enables individuals to interact effectively with their surroundings and perform complex tasks.
This system involves the coordination of various brain regions, including the prefrontal cortex and the parietal lobes, to process and prioritize sensory information, manage tasks, and maintain focus.
In this study, we investigate the intricate mechanisms underpinning the brain's attention system, followed by an exploration within the context of augmented reality (AR) settings.
AR emerges as a viable technological intervention to address the multifaceted challenges faced by individuals with Attention Deficit Hyperactivity Disorder (ADHD).
Given that the primary characteristics of ADHD include difficulties related to inattention, hyperactivity, and impulsivity, AR offers tailor-made solutions specifically designed to mitigate these challenges and enhance cognitive functioning.
On the other hand, if these ADHD-related issues are not adequately addressed, it could lead to a worsening of their condition in AR.
This underscores the importance of employing effective interventions such as AR to support individuals with ADHD in managing their symptoms.
We  examine the attentional mechanisms within AR environments and the sensory processing dynamics prevalent among the ADHD population.
Our objective is to comprehensively address the attentional needs of this population in AR settings and offer a framework for designing cognitively accessible AR applications.

\keywords{Accessibility \and ADHD \and Attention \and Augmented reality \and Neurodiversity}
\end{abstract}
\section{Introduction}
The definition of attention describes it as as a limited resource for information processing~\cite{wickens2014structure}.
Consider an attentionally demanding activity such as driving, a complex task that demands high levels of attention for safety and efficiency.
Driving necessitates the adept handling of multiple cognitive processes: selecting what to focus on, maintaining concentration on critical tasks, managing distractions, and dividing attention among various inputs like traffic conditions, navigation, and vehicle control.
Additionally, the ability to sustain vigilance, especially during prolonged periods of monotonous driving, is crucial~\cite{wickens_engineering_2021}.

Attention acts as a selective filter, enabling us to focus on relevant information while ignoring distractions, thus facilitating cognitive functions such as memory, learning, and decision-making.
However, for individuals ADHD, this filtering process is impaired, leading to difficulties in maintaining focus, managing impulses, and organizing tasks.
ADHD is a prevalent neurobehavioral condition affecting both children and adults worldwide.
It is estimated that up to 5\% of adults and between 6 to 9\% of children and adolescents suffer from ADHD~\cite{adams2009distractibility}, making it a significant concern for public health and individual well-being.
This disorder is characterized by a trio of symptoms: inattention, hyperactivity, and impulsivity, with inattention being a core feature~\cite{10.1145/3010915.3010925} that disrupts social, academic, and occupational functions~\cite{johnson2016antecedents}.
Individuals with ADHD have difficulty staying focused~\cite{volkow2013adult}, following detailed instructions, and organizing tasks.
They may also be easily distracted by irrelevant thoughts or stimuli, and often forget to complete tasks or responsibilities.

AR can intervene and provide support for attention in several innovative ways, including enhancing focus, engagement, and learning through the strategic use of digital augmentations~\cite{keshav2019digital}.
By leveraging visual and auditory cues that blend seamlessly with the real-world environment, AR has the potential to direct the attention of individuals with ADHD towards specific tasks or information, aiding in sustained and selective attention.
Little is known about sustained attention performance of adults with ADHD in AR settings.
This targeted approach is particularly beneficial in mitigating the challenges associated with attention deficit by helping users prioritize sensory and cognitive processing without becoming overwhelmed.
The four principles of attention—select, focus, divide, and sustain—are integral to the effective design and use of AR technologies.
In practice, AR applications balance the presentation of digital augmentations, such as pop-up notifications or virtual objects, with the user's awareness of the real-world environment.
This balance is critical to ensuring safe and effective interaction, as a lapse in attention could result in missed information or failure to notice real-world obstacles.

However, as AR applications becomes widespread, they raises concerns regarding sensory overload, a condition that occurs when an individual's sensory systems are bombarded with more information than they can process, leading to cognitive overload and an overwhelming sensation.
The engagement of multiple sensory modalities in AR—visual, auditory, and sometimes haptic feedback—intensifies the risk of increasing the cognitive load on users.
This potential for sensory overload underscores the importance of careful design in AR technologies, aiming to enrich the user experience without overwhelming their sensory processing capabilities.
Therefore, we tackle ADHD challenges with a design approach to improve attention and daily function, offering key insights for developing inclusive AR solutions for this demographic.

The structure of this paper is organized as follows: Section~\ref{sec:related} reviews related literature, summarizing interventions within the AR domain, including eye-tracking, brain-computer interface (BCIs) technology, and prevalent machine learning algorithms.
Section~\ref{sec:conceptualFramework} presents our conceptual framework, detailing the mechanisms of cognition and perception.
In Section~\ref{sec:attentionGuidingMechanisms}, the challenges faced by individuals with ADHD are discussed, along with corresponding AR interventions and solutions.
Section~\ref{sec:designPorposal} introduces our design proposal, and Section~\ref{sec:conclusion} provides the concluding remarks of the paper.

\section{Related Works}
\label{sec:related}

The adoption of AR technology within ADHD research is garnering significant interest~\cite{bimber2005spatial,carmigniani2011augmented,joseph2023exploring,romero2020self,sweileh2023analysis,vahabzadeh2018improved}.
AR interventions present a promising strategy for ADHD, utilizing immersive experiences to enhance engagement and motivation.
Offering real-time, personalized feedback and adaptable simulations, they facilitate skill training and behavior modification.
This method suits ADHD individuals, potentially improving attention and learning outcomes.
Yet, efficacy varies with symptom severity, personal preferences, and technological flexibility.
Challenges remain in assessing long-term impacts, preventing over-reliance, and overcoming cost and access barriers to ensure equitable use.

\subsection{AR Interventions for ADHD}

AR's potential to improve social interaction, communication skills, and attention in children and adolescents with ADHD has been particularly promising~\cite{sweileh2023analysis}.
This is further supported by Kaimara's exploration, which underscores AR's emerging role as an effective treatment approach, especially when coupled with cutting-edge technologies~\cite{kaimara2022could}.
Additionally, AR introduces a novel dimension of flexibility and practicality by incorporating therapeutic light settings into users' daily routines, offering an innovative avenue for continuous light-based interventions~\cite{10415024}.
Importantly, AR's capacity to enrich the real world with essential information addresses sensory impairments effectively~\cite{bimber2005spatial,carmigniani2011augmented,joseph2023exploring,1667684}.
While its application as an educational tool for children with special needs, including those with ADHD, highlights its interactive and immersive capabilities in facilitating ADHD management in pediatric settings~\cite{vahabzadeh2018improved}.

Extending the scope of AR's impact, innovative approaches have combined AR with ADHD interventions to enhance social interactions~\cite{romero2020self}.
AR's association with positive outcomes in ADHD contexts includes enhancements in behavior, cognitive function, and dual-tasking abilities in children~\cite{shema2019virtual}.
Additionally, AR has been proposed as a valuable tool for enhancing specific cognitive skills, such as reading and spelling, through dedicated literacy programs for children with ADHD~\cite{tosto2021exploring}.
The development of AR serious games (ARGS) aims to train focused and selective attention in children with ADHD~\cite{avila2018towards}.
Research has also explored inattentional blindness~\cite{wickens_engineering_2021} and attentional tunneling~\cite{syiem_impact_2021}, examining cross-modality attention in AR, highlighting its comprehensive benefits in addressing ADHD-related challenges and showcasing its potential as a versatile intervention tool.

Research demonstrates AR's potential in enhancing literacy, specifically improving reading and spelling in children with ADHD~\cite{tosto2021exploring}, by offering tailored educational interventions.
Further, AR's role extends to boosting memory recall in educational contexts, suggesting significant benefits for ADHD-related cognitive processes and information retention~\cite{zhang2019using}.
Game-based AR interventions have also shown to engage children with ADHD effectively, increasing satisfaction and potentially improving quality of life, while also strengthening crucial cognitive aspects~\cite{kim2022adjuvant}.
These studies collectively underscore AR's versatility in addressing diverse challenges associated with ADHD, from literacy gaps to enhancing memory and cognitive engagement.

\subsection{Visual Attention in Eye-tracking}

Eye tracking is a pivotal method for gauging where overt attention is directed, marking observable shifts in focus like gaze changes, in contrast to covert attention's internal shifts without eye movement~\cite{hafed2015vision}.
This distinction aids in comprehending attention's role across various scenarios, including visual perception and spatial cognition.
Leveraging eye-tracking technology, recent studies have probed the dynamics of how attention is oriented and maintained on particular stimuli, such as faces~\cite{leppanen2016using}.
Eye movements, driven by visual-spatial attention, offer a direct measure of the direction and speed of attention in response to stimuli~\cite{zhang2021using}.

In practical terms, eye tracking is invaluable for analyzing attention-related behaviors in ADHD, revealing unique eye movement patterns like diminished fixation on faces and irregular scan paths~\cite{mohammadhasani2020atypical}.
Computerized eye-tracking training has been effective in improving saccadic movements and inhibitory control in ADHD, showcasing its potential for intervention~\cite{lee2020computerized}.
Moreover, in the realm of e-learning, eye tracking supports customized educational strategies for ADHD, enhancing learning outcomes~\cite{khan2022execute}.
This technology's application underscores its significance in both understanding and addressing attention-related challenges, particularly within ADHD populations.

\subsection{Attention in Brain-Computer Interfaces}
Attention plays a pivotal role in Brain-Computer Interfaces (BCIs) research, influencing both the performance and applicability of BCIs across different domains.
Electroencephalography (EEG), a non-invasive technique for monitoring brain activity, is instrumental in evaluating attention, particularly within AR environments.
Studies have explored the use of EEG and eye tracking to distinguish between focused and divided attention in individuals wearing AR headsets, showing the capability of these technologies to monitor attention levels~\cite{kosmyna2021assessing}.
Moreover, EEG has revealed increased engagement and attention during AR tasks, marked by enhanced brain activity~\cite{lim2023development}.
In the context of ADHD, multimedia interventions have been assessed through EEG, demonstrating improved attentional focus, as indicated by decreased theta wave activity~\cite{10.1145/2093698.2093714}.
Additionally, ethical considerations in BCI research are gaining attention, highlighting the importance of thorough evaluation and refinement of BCIs to enhance performance while addressing ethical implications.

\subsection{Attention in Machine Learning}

Machine learning (ML) algorithms are gaining traction for their role in AR interventions for ADHD, showcasing their capability in identifying and classifying the disorder through advanced neuroimaging techniques like MRI~\cite{ghiassian2016using,eslami2021machine}.
Research employing ML on extensive ADHD datasets has proven its effectiveness in the accurate identification and classification of ADHD, further supported by the achievement of detection accuracies up to 85.45\%~\cite{eslami2021machine,10035012}.
This highlights ML's significant potential in refining ADHD diagnosis and treatment strategies.

ML algorithms have been crafted for distinguishing ADHD from non-ADHD children using EEG signals, showcasing ML's broad applicability in ADHD diagnosis~\cite{maniruzzaman2022efficient}.
Further, ML aids in differentiating ADHD subtypes, assisting clinicians in timely patient identification~\cite{lin2023distinguishing}.
Its application extends to accurately distinguishing between autism and ADHD, demonstrating ML's versatility in identifying neurodevelopmental disorders~\cite{duda2016use}.
Moreover, ML's role in ADHD classification has evolved towards creating diagnostic tools incorporating VR and deep learning, highlighting innovative approaches in ADHD diagnostics~\cite{tosto2021exploring,wiguna2020four}.
Additionally, a Gaussian SVM-based ML model focusing on linear EEG features has proven effective in ADHD detection in clinical settings, demonstrating high accuracy and underscoring ML's role in early diagnosis and treatment.
This approach emphasizes ML's capacity to improve diagnostic precision while minimizing computational demands through extensive data sets and stringent validation processes~\cite{alim2023automatic}.

A cutting-edge EEG signal analysis method enhances ADHD detection by integrating Variational Mode Decomposition (VMD) with Hilbert Transform (HT).
This approach refines the extraction and analysis of EEG rhythms—delta, theta, alpha, beta, and gamma—improving accuracy.
By calculating entropy-based features from these rhythms, the technique assesses its efficiency in differentiating ADHD from non-ADHD subjects.
Notably, the Extreme Learning Machine (ELM) classifier exhibits exceptional sensitivity, accuracy, and specificity in this context.
The innovative VHERS model, as a result, stands out for its automatic ADHD detection capabilities, marking a substantial progress over previous methods and underscoring its applicability in real-time scenarios~\cite{khare2022vhers}.

\section{Conceptual Framework}
\label{sec:conceptualFramework}

Designing a context-aware AR system for ADHD necessitates a comprehensive integration of sensor technology, machine learning, and multimodal engagement strategies to dynamically adapt to an individual's environment and activity patterns.
This approach ensures the system can effectively direct the user's attention towards relevant physical and virtual objects, enhancing engagement and seamlessly incorporating technology into daily life as suggested by Biocca et al.~\cite{10.1145/1124772.1124939}.

By leveraging advanced sensors and machine learning algorithms, the system gains a deep understanding of the user's context, including movement and environmental conditions, to select the most suitable modality—whether visual, auditory, or tactile—for engaging the user.
This adaptive mechanism, underscored by a multidisciplinary approach that combines user experience design with robust data security, aims to tailor AR experiences to the unique needs of individuals with ADHD, minimizing sensory overload and enhancing cognitive functions.

Sensor integration, including accelerometers, gyroscopes, and GPS, provides valuable data on user orientation, motion patterns, and precise location tracking.
Additionally, computer vision techniques through camera feeds offer insights into user movements, particularly in indoor settings where GPS may be less effective.
Understanding the user's environment extends to deploying environmental sensors to assess ambient conditions such as light, temperature, and noise levels.
This data, combined with location-based services and advanced mapping technologies, offers a comprehensive understanding of the user's context.
Machine learning models trained on this diverse data can discern intricate patterns, enhancing the system's contextual awareness.

The importance of selecting the most appropriate modality for engaging user attention through a nuanced analysis of preferences, contextual relevance, and effectiveness of past interactions, facilitated by adaptive algorithms.
These algorithms adjust to user feedback and changing conditions, prioritizing auditory notifications in visually demanding situations and haptic feedback in noisy environments, ensuring responsiveness to the dynamic nature of human activity.
The integration of machine learning and artificial intelligence within a robust architectural framework is crucial for processing real-time data, adapting to shifts, and learning from user behavior, while maintaining stringent data security measures to protect privacy.
Additionally, the text highlights the impact of ADHD on AR cognitive and neuroergonomic aspects, advocating for adaptive AR designs that minimize sensory overload risks and enhance cognitive functions, emphasizing the need for research into sensory overload, optimal exposure durations, and tailored AR experiences for learning and problem-solving.

The development of context-aware AR systems necessitates an in-depth understanding of human information processing, particularly how attention and cognition interact within these environments.
Recognizing how individuals process, interact with, and respond to information can significantly refine AR technology design, making these systems more intuitive, effective, and cognitively compatible.
This exploration into the nuances of attention—its influence on memory, multi-tasking, and decision-making—provides essential insights into optimizing AR experiences.
By dissecting the complex dynamics of attentional mechanisms and their impact on cognitive functions, we aim to enhance AR technologies in ways that seamlessly align with human cognitive processes, thereby enriching user interaction and overall experience.

\begin{figure}
\centering
\includegraphics[width=0.8\textwidth]{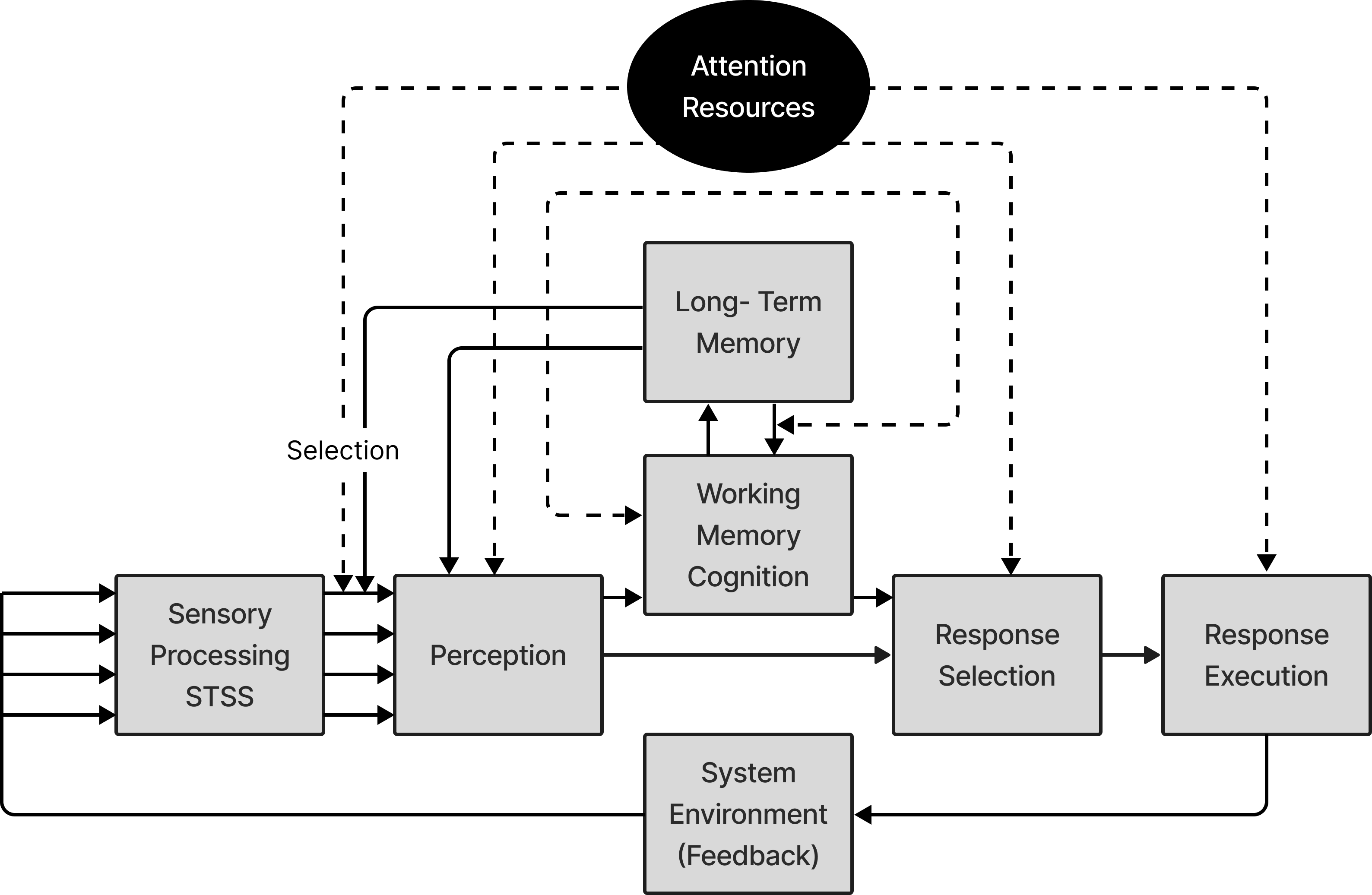}
\caption{A model of human-information processing stages~\cite{wickens_engineering_2021}.}
\label{fig1}
\end{figure}

\subsection{Human Information Processing}
According to Wicken's human information processing model~\cite{wickens_engineering_2021} as depicted in Figure~\ref{fig1}, sensory information is initially gathered through the Short-Term Sensory Store (STSS), where an abundance of environmental stimuli is received through our senses.
The selection process then filters this information, guided by attention resources, to determine which stimuli will proceed to the next stage of processing.
Once selected, this sensory input is channeled into perception, where it is interpreted, organized, and transformed into a coherent representation of the environment.
Attention resources play a pivotal role throughout this process, as they allocate the cognitive bandwidth necessary to manage the flow and processing intensity of the information.
Simultaneously, the model illustrates a bidirectional interaction with memory.
Long-term memory provides a repository of past experiences and knowledge that can influence perception and decision-making processes.
Conversely, new information can be consolidated into long-term memory for future use.
Working memory serves as an active workspace where information is temporarily held and manipulated, facilitating cognitive functions such as reasoning, comprehension, and learning.
Decision-making culminates in the response selection stage, where, based on the interplay of perception, memory, and cognitive evaluation, a suitable response to the stimuli is chosen.
Following this, response execution occurs, wherein the selected action is carried out, whether it be a physical movement, verbal output, or another form of observable behavior.
The model closes the loop with a system environment feedback component, which implies that the consequences of the executed response are observed and assessed, influencing subsequent sensory processing.
This feedback mechanism ensures that cognitive processing is adaptive, allowing for the refinement of responses based on the success or failure of previous interactions with the environment.
Thus, the model encapsulates a cyclical and interactive framework, demonstrating the complexity and dynamism of human information processing.

Compared to typical individuals, those with ADHD may show unique differences in their cognitive processing due to the disorder's fundamental issues related to attention, hyperactivity, and impulsivity.
At the sensory processing stage, those with ADHD might experience an overload of sensory information, as their STSS may not filter stimuli as effectively as in non-ADHD individuals.
This can lead to a sensation of being overwhelmed by sensory data, making it difficult to distinguish between relevant and irrelevant information.
When it comes to selection, the attentional deficits characteristic of ADHD make it challenging to maintain focus on specific stimuli, causing frequent distractions and shifts in attention.
This inconsistency in selection directly impacts perception, as the interpretation of sensory information can become erratic.

Attention resources in ADHD are often compromised, which manifests in a reduced ability to allocate cognitive bandwidth where it's most needed, affecting the entire processing system's efficiency.
The interaction with memory systems is also affected; while long-term memory storage might remain intact, the retrieval process can be hampered due to the working memory and attentional issues.
Working memory in ADHD individuals tends to be less robust, leading to difficulties in holding and manipulating information, which is vital for carrying out cognitive tasks that require planning and reasoning.
Impulsivity, a hallmark of ADHD, can significantly influence response selection, leading to rapid decision-making often without full consideration of the consequences.
This impulsiveness, coupled with potential hyperactivity, results in response execution that may be hasty or poorly coordinated.
The actions taken might be less consistent and more variable than those of individuals without ADHD.

Ultimately, the system environment feedback, which closes the loop in cognitive processing, may not be as effective in those with ADHD.
Due to the difficulties in sustaining attention and working with memory, learning from past experiences and adjusting subsequent behaviors can be challenging.
Feedback may not lead to the same degree of behavioral adjustment as it might in individuals without ADHD, leading to a cycle that may not self-correct as efficiently.
Interventions for ADHD typically focus on enhancing the efficiency of these cognitive processes, aiming to bolster attentional control, decrease impulsivity, and improve working memory capabilities.

\subsection{Attention and Working Memory}

Working memory (WM) and attention are crucial cognitive functions that are often affected in individuals with ADHD.
Understanding how these processes occur in ADHD involves recognizing the challenges posed by the disorder and their impact on cognitive performance.
WM is fundamentally about the top-down, active manipulation of data stored in short-term memory, a process central to cognitive functioning~\cite{jones2023privacy,kofler2020working}.
This cognitive system, responsible for the temporary storage and manipulation of information, is tightly interwoven with attention.
The consensus in the scientific community is strong regarding the intimate linkage between WM and attention, highlighting their mutual dependence for efficient cognitive processing~\cite{awh1998rehearsal,oberauer2019working}.
Attention not only plays a pivotal role in the entry and maintenance of information within working memory but also in the manipulation of this data, underscoring the symbiotic relationship between these two processes~\cite{eriksson2015neurocognitive}.

This interdependence is further elucidated through the dynamics of bottom-up and top-down processes; attention facilitates the initial transfer of information into working memory while also being subject to control by the contents of working memory itself, showcasing a bidirectional interaction~\cite{10.1016/j.tics.2012.10.003,10.1177/0956797612439068}.
Moreover, the capacity of multiple WM representations to concurrently direct attention emphasizes the complex, simultaneous control of attention by various elements within working memory~\cite{10.1016/j.neuron.2006.01.021}.

Traditionally viewed as a one-way process where attention filters and allows only relevant information into short-term processing stores, the relationship between WM and selective attention is now recognized to be more nuanced~\cite{10.3758/s13414-019-01721-8}.
The debate over the extent to which working memory influences attentional processes points to a more intricate and reciprocal interaction than previously thought, highlighting the sophistication of cognitive mechanisms at play~\cite{10.1037/a0016452}.
WM deficits have been associated with the primary behavioral symptoms of ADHD, including inattention, hyperactivity, impulsivity, and the overall severity of the disorder~\cite{kofler2018episodic}.

\subsection{Attention and Decision Making}

Attention and decision-making are complex cognitive processes that are closely intertwined.
Research has shown that attention plays a crucial role in guiding decision-making processes.
For instance, studies have demonstrated a triple dissociation of attention and decision computations across the prefrontal cortex (PFC)~\cite{hunt2018triple}.
This suggests that different subregions of the PFC are involved in attention-guided decision-making tasks.
Furthermore, covert attention has been found to lead to faster and more accurate decision-making, highlighting the importance of attention in decision processes~\cite{perkovic2023covert}.
Additionally, the visual environment has been identified as a fundamental aspect of everyday decisions, with attention being a key factor in this process~\cite{braver2012variable}.
However, impulsive and risky decision-making are associated with ADHD~\cite{dekkers2022impulsive}.
Research has shown that individuals with ADHD exhibit suboptimal and inconsistent temporal decision-making, as well as risky decision-making, indicating a link between ADHD symptoms and decision-making inconsistencies~\cite{gabrieli2022symptoms}.
The inclination towards risky decision-making in the ADHD population include difficulties with executive functioning, such as planning, inhibition control, and working memory, all of which are crucial for thoughtful decision-making.
The exploration of risky decision-making within the ADHD demographic has yet to be thoroughly investigated.
This gap in research presents an opportunity for scholars to delve deeper into the safety aspects of AR and its implications for individuals with ADHD.

\subsection{Multi Task Attention}

Attention and multitasking are critical cognitive functions that influence academic outcomes, particularly in individuals with ADHD, who often face challenges in managing multiple tasks simultaneously due to difficulties in sustaining attention and handling competing cognitive demands~\cite{kofler2018episodic}.
Digital interventions targeting ADHD have demonstrated improvements in attention, inhibition, and working memory, suggesting that such approaches can mitigate multitasking deficits in affected individuals~\cite{kollins2015novel}.
Despite these advancements, the impact of technologies like VR on cognitive overload and attention spans in ADHD remains unclear, with research including that by Kaimara et al., indicating the need for further investigation into the relationship between ADHD and immersive technologies like AR and VR~\cite{kaimara2022could}.
This highlights the potential of targeted interventions while also acknowledging the complexities of technology's effects on cognitive functions in ADHD.

\section{Attention Guiding Mechanisms}
\label{sec:attentionGuidingMechanisms}

Researchers have not yet agreed on how to best guide attention in immersive environments, though AR serious games and neurofeedback stand out as effective techniques.
This opens a door to investigate enhancing attention via innovative methods like 3D UIs and adaptive context-aware technologies.
This gap presents an opportunity to explore how attention can be enhanced through innovative approaches such as 3D user interfaces (UIs) and context-aware technologies.
based on our literature review, several key features and considerations are paramount to ensure its effectiveness and accessibility, as depicted in Table~\ref{tab:my_label}.
It is important to mention that, after conducting our research on a large scale and validating our proof of concept, this table will be updated to become more comprehensive.

\begin{longtable}[c]{| m{0.45\linewidth} | m{0.55\linewidth} |}
\caption{Challenges associated with ADHD and corresponding AR interventions.}
\label{tab:my_label}\\

\hline
\textbf{ADHD Challenges} & \textbf{AR Interventions and Solutions} \\
\hline
\endfirsthead

\multicolumn{2}{c}%
{{\bfseries Table \thetable\ continued from previous page}} \\
\hline
\textbf{ ADHD Challenges} & \textbf{AR Interventions and Solutions} \\
\hline
\endhead

\hline
\endfoot

\hline

\endlastfoot

Difficulty with Sustained Attention & Attention-Enhancing Tasks or Spatial Cues: Interactive tasks that progressively increase in complexity, along with spatial cues, to cultivate longer attention spans\\
\hline
Impaired Working Memory & Enhanced Cues and Interactive Feedback: Utilize visual and auditory cues, concise and digestible interface design, real-time feedback, and strategic repetition to aid memory retention and processing \\
\hline
Increased Distractibility & Customizable Sensory Environments: Offer control over sensory inputs to minimize distractions, ensuring focus remains on relevant tasks and stimuli \\
\hline
Executive Function Challenges & Gamified Learning Experiences: Incorporate gamification to enhance cognitive flexibility, planning, and decision-making through engaging and educational content \\
\hline
Difficulty with Spatial Orientation & Adaptive Navigation Aids: Utilize individual performance metrics to tailor navigation aids in 3D spaces, enhancing spatial orientation and reducing cognitive load \\
\hline
Overreliance on Multitasking & Real-Time Feedback Systems: Provide immediate feedback to manage multitasking effectively, reinforcing task completion and smooth transitions \\
\hline
Challenges with UI and Interaction Design & Intuitive Gamified Interfaces: Develop accessible, intuitive UIs using gamification to simplify navigation and interaction, reducing cognitive strain for ADHD users \\
\hline
Motivational Fluctuations & Personalized Engagement Strategies: Employ data analytics to customize tasks and rewards according to user motivation levels, fostering sustained engagement and interest \\
\hline
Processing Speed Variabilities & Pace-Adjustable Sensory Inputs: Allow users to adjust the pace of sensory inputs, accommodating different processing speeds and reducing the risk of overwhelm\\
\end{longtable}

Several challenges inherent to ADHD can impact the experience within AR.
Sustained attention is often a hurdle, as AR environments demand extended focus, which can be taxing for ADHD users, leading to faster cognitive fatigue.
WM impairments, common in ADHD, can make it challenging to retain and manipulate information in real-time, thereby increasing cognitive load and complicating task performance.
In the context of AR, addressing working memory deficits in individuals with ADHD is crucial for optimizing their AR experience.
AR applications can be designed to accommodate working memory impairments by providing visual and auditory cues to aid in memory retention and retrieval.
Additionally, the design of AR interfaces should consider the limited working memory capacity of individuals with ADHD, ensuring that the information presented is concise and easily digestible.
Furthermore, incorporating interactive elements and real-time feedback in AR experiences can help individuals with ADHD engage their working memory more effectively.

The rich sensory nature of AR can also introduce numerous distractions; irrelevant stimuli can divert attention from the task at hand, impeding progress.
Moreover, executive function challenges in planning, decision-making, and cognitive flexibility are significant obstacles when AR simulations present complex scenarios.
Spatial orientation in three-dimensional spaces is another area of difficulty, especially when environments are intricate or navigation aids are lacking.
An overreliance on multitasking within AR can heighten cognitive overload and impair task efficiency due to the frequent need to shift attention.
User interface and interaction design pose additional challenges; non-intuitive or complicated interfaces can be especially burdensome for individuals with ADHD who may find it difficult to learn and recall navigational procedures.

Customizable sensory inputs allow users to tailor their sensory experience to avoid overstimulation or understimulation, which is critical for those prone to sensory processing challenges.
The environment incorporates attention-enhancing tasks that become progressively more complex, helping users to expand their attention spans and cognitive processing abilities gradually.

An immediate feedback system is a cornerstone of this AR environment, offering users real-time responses that are essential for reinforcing learning outcomes and encouraging positive behavior, thus fostering motivation and a sense of achievement.
The content within the AR is both educational and engaging; it leverages gamification techniques to maintain user interest and facilitate skill development and learning.
Moreover, the use of data analytics is instrumental in personalizing the experience; by analyzing individual performance metrics, the environment adjusts task difficulty and types, thereby delivering a tailored and effective learning journey.
These integrated features collectively aim to present an innovative educational tool designed specifically for the ADHD community, with the potential to markedly improve attentional control and learning outcomes.

Motivational fluctuations, often associated with ADHD, can affect user engagement with AR tasks, particularly if the environment lacks immediate feedback or rewards, potentially leading to reduced involvement and higher cognitive strain.
Lastly, variabilities in processing speed may influence the ability of ADHD users to swiftly comprehend and respond to stimuli within AR simulations, which can result in feelings of overwhelm or frustration.
These challenges necessitate careful design and consideration to ensure AR environments are accessible and supportive for users with ADHD.

\begin{figure}
\centering
\includegraphics[width=0.8\textwidth]{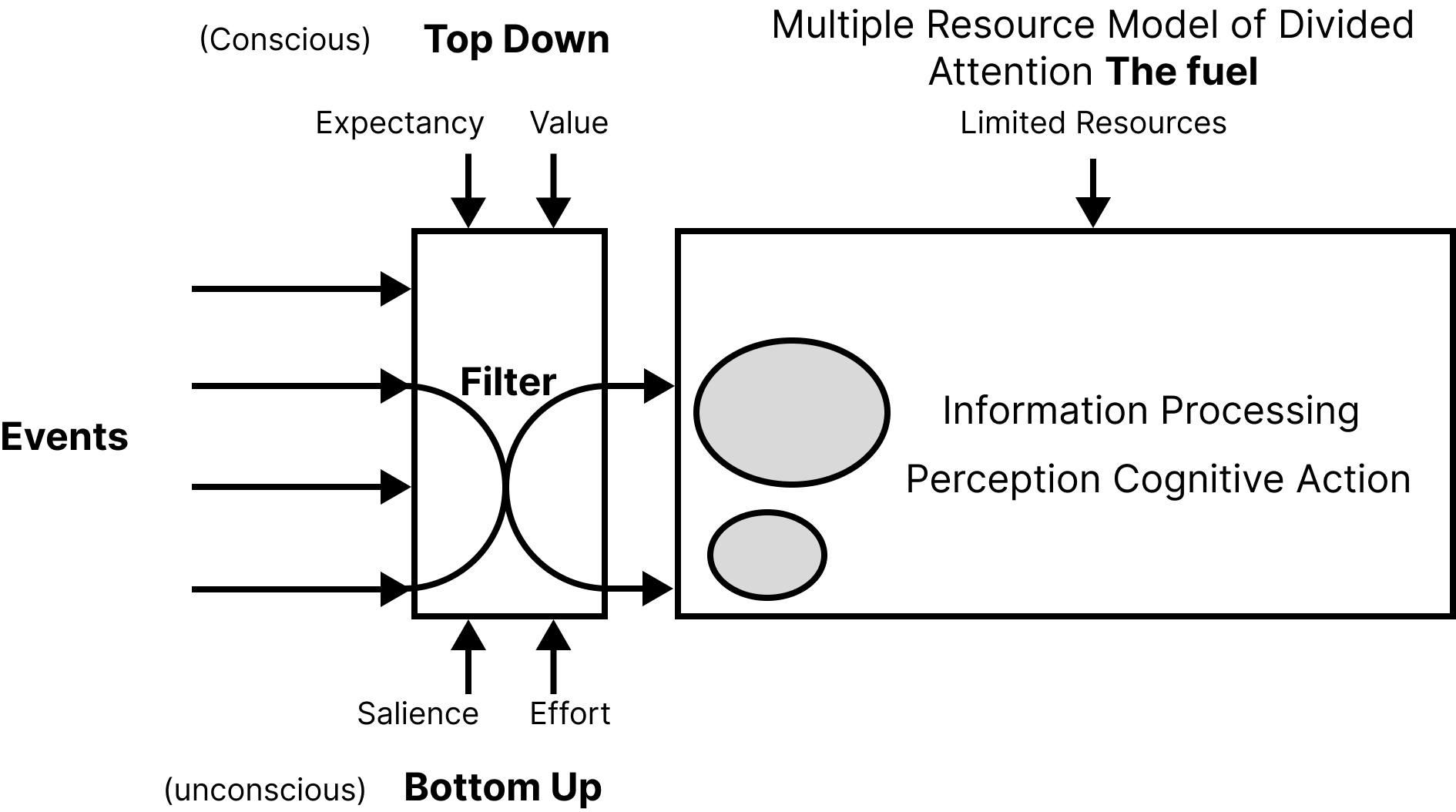}
\caption{SEEV model in design~\cite{wickens_engineering_2021}.} 
\label{fig2}
\end{figure}

\section{Design}
\label{sec:designPorposal}

\begin{figure}
\centering
\includegraphics[width=0.8\textwidth]{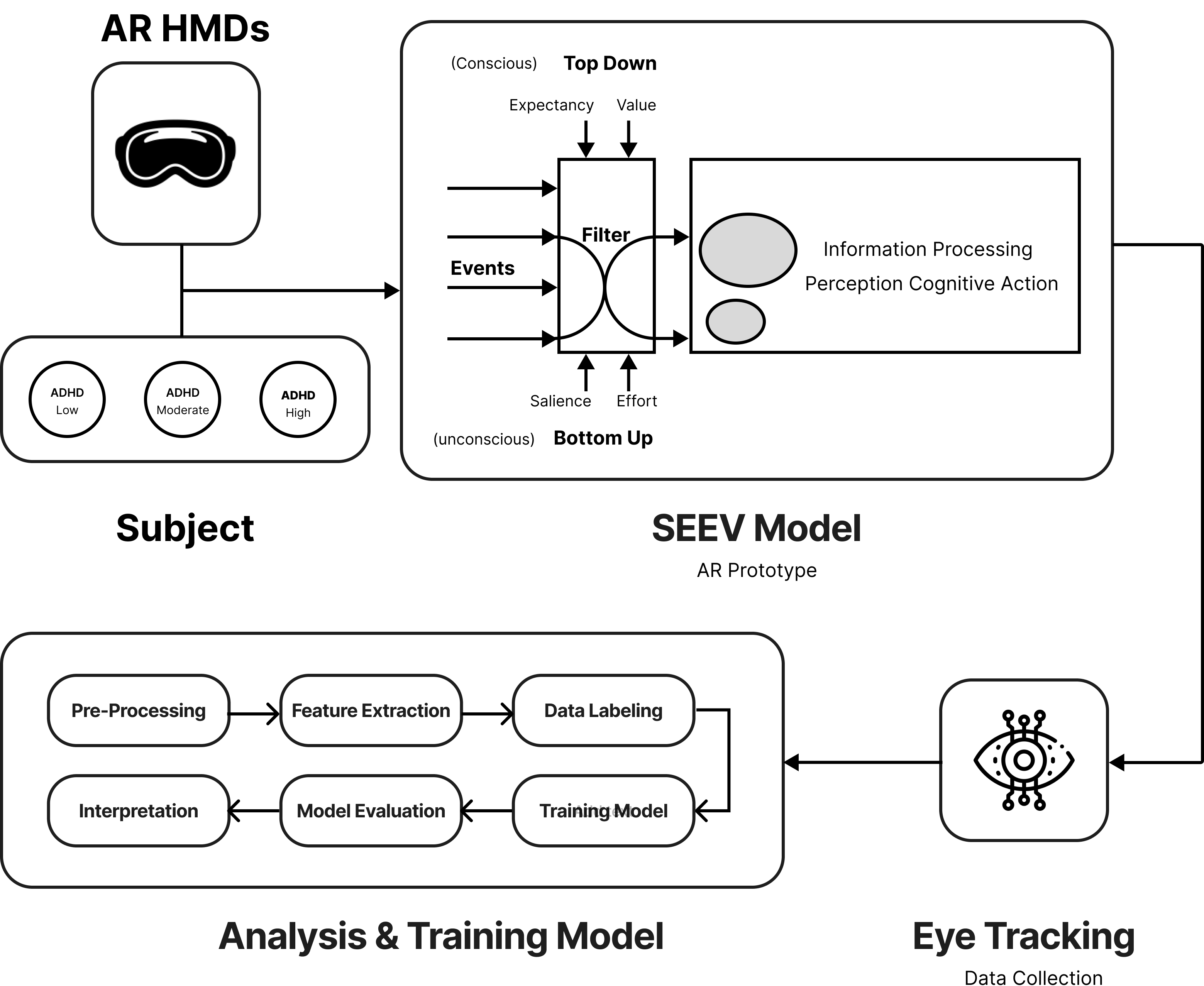}
\caption{Conceptual framework.} 
\label{fig3}
\end{figure}

Our study introduces a novel design concept that applies the SEEV model within an AR setting, specifically tailored to meet the varied requirements of individuals with ADHD.
The SEEV model, recognized for its predictive capabilities in visual attentionc~\cite{wickens_engineering_2021}, incorporates four key factors—Salience, Effort, Expectancy, and Value—that are instrumental in the design of effective 3D UIs.
By assessing the visual prominence of UI elements (Salience), the physical or cognitive exertion required to interact with them (Effort), the likelihood of encountering important information (Expectancy), and the perceived benefit of the interaction (Value), the model provides a comprehensive framework for managing and directing user attention.
In Figure~\ref{fig3}, a pipeline of our work depicted our design system that begins with the application of the SEEV model to enhance situational awareness within an AR domain.
Following the SEEV model, we implement our prototype in  Apple Vision Pro headset, equipped with built-in eye-tracking technology used to collect eye tracking data to analyze complex patterns of fixations and saccades in relation to the visual stimuli and tasks can help distinguish between top-down and bottom-up attention influences.

AR prototype needs to accommodate a spectrum of ADHD types —categorized as low, medium, and high.
This differentiation is essential to tailor the AR experience to the unique attentional capacities of each group.
To this end, the prototype encompasses a triad of dual-task scenarios, each crafted to evaluate multitasking proficiency through the deployment of varied sensory modalities.
The first scenario focuses on information retrieval, wherein participants are engaged in auditory information processing—such as listening to a lecture—while concurrently utilizing AR technology to search for specific information.
The second scenario encompasses dynamic navigation, challenging participants to traverse a physical milieu whilst managing digital interactions, exemplified by holding a conversation or responding to virtual notifications through the AR interface.
The third scenario is a collaborative endeavor, wherein participants engage in a live group discussion while processing textual information, such as reading an email, via the AR display.

Following task identification, eye-tracking technology is employed to collect data from participants engaged in these tasks.
A particular emphasis is placed on including a substantial cohort of individuals diagnosed with ADHD, alongside a control group, to facilitate a comparative analysis.
The setup for data collection is standardized, ensuring consistency across various tasks and participant experiences.
This phase is critical for capturing the nuanced eye movement patterns that signify attentional engagement or distraction.

Prior to analysis, the collected eye-tracking data undergo a rigorous preprocessing phase.
This includes the removal of noise, correction of data drifts, and addressing missing data points, ensuring the integrity of the dataset for feature extraction.
Relevant features, such as fixation duration,  saccade movements, and blink rate, are identified and extracted.
These features are posited to be indicative of focused attention and are thus critical for the subsequent classification phase.

The core of our methodology involves employing Support Vector Machines (SVM) for the classification of preprocessed eye-tracking data.
This process begins with a careful selection of features that robustly indicate attention patterns.
The SVM model is then trained with a subset of this data, with model optimization undertaken to enhance accuracy and reliability through hyperparameter tuning.
Validation of the model is conducted using a separate dataset, with performance evaluated through standard metrics such as accuracy, precision, recall, and F1 score.

The SVM algorithm is used from the scikit-learn library, a widely utilized machine learning toolkit in Python, to develop a classification model for our eye tracking data.
The initial phase of our approach involved preparing the dataset, which was followed by the training of an SVM model to categorize each coordinate pair (x, y) into one of three categories: Area 1, Area 2, or neither.
The feature set comprised the (x, y) coordinates, utilized to ascertain the focal area of the user.
To effectively accomplish this, it was imperative to delineate the boundaries for Areas 1 and 2.
For the sake of simplicity, we postulated that these areas could be approximated by rectangles, although more complex shapes could be considered depending on specific requirements.
Additionally, certain data points, referred to as ``pert area'' points, were identified as either outliers or points of specific interest, necessitating further clarification regarding their role in the analysis of eye tracking.

During the data preparation phase, we assigned labels to the data contingent upon their designated areas: points within the confines of Area 1 received a label of 1, those within Area 2 were labeled 2, and points outside both areas were labeled 0.
In terms of feature selection, we considered the x and y coordinates as primary features.
The duration of gaze, represented by the 'time' attribute, was also evaluated as a potential feature, given its possible influence on classification outcomes.
For model training, we utilized scikit-learn's Support Vector Classification (SVC), opting for the Radial Basis Function (RBF) kernel.
The RBF kernel was selected due to its proven efficacy in managing non-linear data separations and intricate patterns, a common characteristic in eye tracking data where the subjects of interest, such as text and faces, do not manifest as linearly separable clusters in the feature space.
Parameter selection was guided by existing literature, setting the regularization parameter C to a range of \{26.5, 26.75, 27, 27.25, 27.5, 27.75, 28\}.
This decision was based on a study that identified optimal ranges for this parameter~\cite{budiman2019svm}.
Similarly, the gamma parameter for the 'rbf' kernel was set to $\gamma = \{2^{-14.5}, 2^{-14.75}, 2^{-15}, 2^{-15.25}, 2^{-15.5}, 2^{-15.75}, 2^{-16}\}$, with the highest classification accuracy achieved at $C = 27$ and $\gamma=2^{-15}$, indicating that such methodological approaches could be beneficial for optimizing parameters in eye tracking data classification.

In the final stage of model prediction and analysis, we leveraged the trained model to determine the focal area for each time point, allowing us to quantify the total duration spent focusing on each area, thus facilitating a comprehensive understanding of user engagement with the designated areas.
Analysis of the SVM classification results enables the identification of distinct patterns of visual fixation and attention distribution, particularly among individuals with ADHD.
These patterns offer unprecedented insights into the attentional dynamics facilitated by AR technology, highlighting differences and similarities in attention engagement compared to the control group.
The findings of this research not only contribute to the academic understanding of ADHD in the context of emerging technologies but also offer practical implications for the design of AR experiences.
Recommendations are provided for creating AR environments that are cognizant of the attentional needs of individuals with ADHD, potentially enhancing their engagement and learning outcomes.
Furthermore, this study opens avenues for future research, suggesting the exploration of alternative machine learning models and the investigation of different cognitive tasks within AR settings.

Our pilot study has yielded promising initial findings regarding the use of AR technologies to aid individuals with ADHD in managing their attention.
The study engaged a select group of individuals from the ADHD community, employing a prototype to demonstrate the potential of AR in facilitating attention management.
Acknowledging the constraints imposed by the small sample size, we plan to expand our research to include a more diverse group of participants.
The subsequent phase of the study will incorporate  statistical analysis methods such as the t-Test and ANOVA, in conjunction with NASA Task Load Index (NASA-TLX), System Usability Scale (SUS), and User Experience Questionnaire (EUQ) questionnaires, aiming to compile a comprehensive dataset.
This enriched dataset is expected to yield more precise and generalizable insights, deepening our understanding of how different ADHD groups allocate visual attention in AR environments.
The findings from this research are projected to be instrumental in shaping the development of AR experiences that are finely tuned to meet the varied cognitive needs of users.

Additionally, the investigation aims to unearth patterns in eye movement and fixation data, which are instrumental in the creation of AR applications that are finely tuned to the attentional requirements of the ADHD population.
By doing so, we strive to amplify their proficiency in managing attentional resources amidst multifaceted scenarios.
The research identifies a conspicuous void in the current landscape—the absence of a design model adept at pinpointing the most conducive layout within immersive environments.
Addressing this need is imperative, given that cognitive processes exhibit substantial variability across individuals.
The quantity of information presented and its strategic placement are pivotal considerations that directly impact user satisfaction and efficacy.

\section{Conclusion}
\label{sec:conclusion}

We introduced a novel approach to enhance situational awareness in AR applications, particularly for individuals with ADHD.
We identified a few gaps in current research when SEEV model have not been explored yet in AR domain for ADHD population.
We proposed a comprehensive data-driven strategy, employing advanced eye-tracking technology and machine learning techniques, to scrutinize visual attention mechanisms.
By conducting this research with a larger and diverse ADHD population, the study anticipates a robust foundation for further development in AR technologies.
This research not only addresses a significant gap by exploring the nuances of attention management among individuals with ADHD in AR settings but also sets the stage for future developments in AR technology applications for educational, therapeutic, and recreational purposes for individuals with diverse cognitive profiles.

%
%
%

\bibliographystyle{splncs04}
\bibliography{Bibliography}
\end{document}